\documentclass[letterpaper]{article} 
\usepackage{aaai24}  
\usepackage{times}  
\usepackage{helvet}  
\usepackage{courier}  
\usepackage[hyphens]{url}  
\usepackage{graphicx} 
\usepackage{natbib}  
\usepackage{caption} 
\usepackage{algorithm}
\usepackage[noend]{algpseudocode}

\usepackage{newfloat}
\usepackage{listings}
\DeclareCaptionStyle{ruled}{labelfont=normalfont,labelsep=colon,strut=off} 
\floatstyle{ruled}
\newfloat{listing}{tb}{lst}{}
\floatname{listing}{Listing}

\usepackage{hhline}
\usepackage{multirow}
\usepackage{placeins}
\usepackage{amssymb}
\usepackage{amsmath}

\usepackage{subcaption}

\newcommand{\pluseqB}{\mathrel{+}=}
\newcommand{\minuseqB}{\mathrel{-}=}

\title{Improving Learnt Local MAPF Policies with Heuristic Search}
\author{
    Rishi Veerapaneni\equalcontrib\textsuperscript{\rm 1},
    Qian Wang\equalcontrib\textsuperscript{\rm 2},
    Kevin Ren\equalcontrib\textsuperscript{\rm 1},
    Arthur Jakobsson\equalcontrib\textsuperscript{\rm 1},\\
    Jiaoyang Li\textsuperscript{\rm 1}, 
    Maxim Likhachev\textsuperscript{\rm 1}
}
\affiliations{
    \textsuperscript{\rm 1}Carnegie Mellon University\\
    \textsuperscript{\rm 2}University of Southern California\\
    \{rveerapa,kevinren,ajakobss\}@andrew.cmu.edu, pwang649@usc.edu, \{jiaoyanl,maxim\}@cs.cmu.edu

}

\begin{document}

\maketitle

\begin{abstract}
Multi-agent path finding (MAPF) is the problem of finding collision-free paths for a team of agents to reach their goal locations. 
State-of-the-art classical MAPF solvers typically employ heuristic search to find solutions for hundreds of agents but are typically centralized and can struggle to scale when run with short timeouts.
Machine learning (ML) approaches that learn policies for each agent are appealing as these could enable decentralized systems and scale well while maintaining good solution quality. Current ML approaches to MAPF have proposed methods that have started to scratch the surface of this potential. However, state-of-the-art ML approaches produce ``local" policies that only plan for a single timestep and have poor success rates and scalability. Our main idea is that we can improve a ML local policy by using heuristic search methods on the output probability distribution to resolve deadlocks and enable full horizon planning. We show several model-agnostic ways to use heuristic search with learnt policies that significantly improve the policies' success rates and scalability. 
To our best knowledge, we demonstrate the first time ML-based MAPF approaches have scaled to high congestion scenarios (e.g. 20\% agent density).

\end{abstract}

\section{Introduction}

The increasing availability of robotic hardware has increased the importance of planning for robotic teams instead of individual robots. These multi-agent robotic systems will enable a multitude of capabilities like rapid search and rescue, exploration on Mars, and efficient warehouse management which may require hundreds of robots to move boxes. Multi-agent systems are appealing as each robot can be cheap and relatively simple while the entire system is scalable and can achieve complex goals.

A fundamental problem with multiple robots is determining how each robot should move. Without careful consideration, robots could be stuck in a deadlock where multiple robots prevent each other from progressing toward their goals. Multi-agent path finding (MAPF) research focuses on developing algorithms for finding collision-free paths for a team of agents to reach their target locations in an efficient and safe manner. 
Although there are several possible approaches to tackle MAPF, the vast majority of these MAPF methods are heuristic search-based methods. These methods have optimality or bounded suboptimality solution guarantees and can solve long-horizon MAPF problems. However, these methods typically trade off solution quality with compute time and require a centralized planner.

Machine learning (ML) approaches that learn policies for each agent are appealing as these could enable decentralized systems that scale well while maintaining good solution quality. Current ML-based MAPF work methods are starting to scratch the surface of this potential. State-of-the-art ML MAPF approaches learn ``local" policies that take in local observations and output a 1-step action distribution.
However, ML in general struggles with long-horizon planning and low-error situations; both are critical in MAPF which requires long-horizon planning across many agents with very little room for errors which can cause collisions or deadlock. Thus, recent ML-based MAPF works have been proposed for gridworlds but currently do not reach the standards of state-of-the-art heuristic MAPF solvers. 

\begin{figure}[t]
    \centering
    \includegraphics[width=0.45\textwidth]{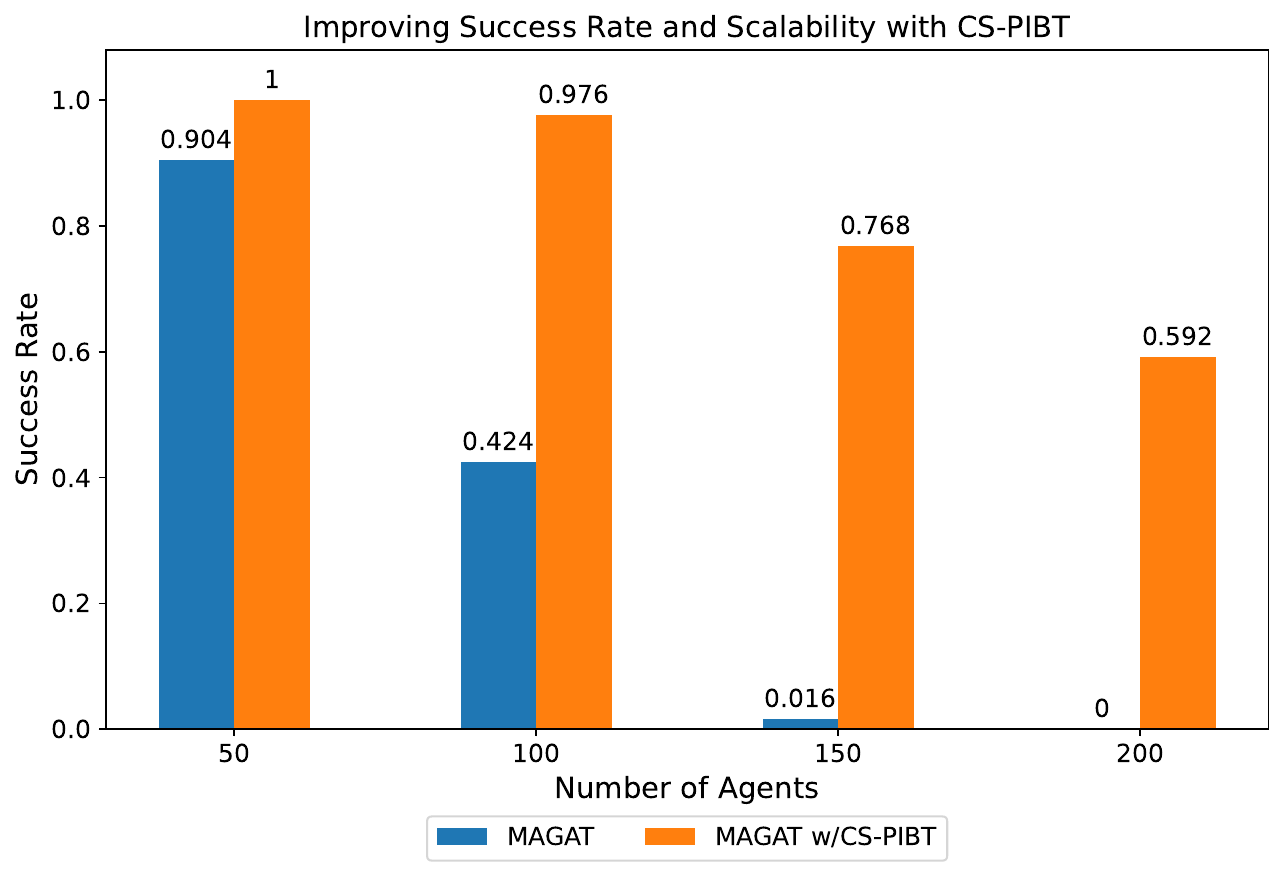}
    \caption{We compare running MAGAT \cite{li2021magat} with its default collision shielding (blue) vs running MAGAT with our PIBT collision shielding (orange). MAGAT is a learnt local policy that predicts a one-step policy per agent that could lead to collisions and therefore requires collision shielding to prevent collisions. PIBT \cite{pibt} is a heuristic search technique for solving MAPF. We see that using the exact same learnt model with our PIBT-based collision shielding improves performance and scalability without any additional training or information.}
    \label{fig:boosting-success-rate}
    \vspace*{-1em}
\end{figure}

Our main idea is that we can boost learnt MAPF approaches by using them with heuristic search methods. When learnt policies predict actions that cause agents to collide, current ML approaches use a naive ``collision shield" that usually replaces collisions with deadlock (see Figure \ref{fig:main-figure}). We instead show how we can use PIBT \cite{pibt} (a heuristic search method) as a smart collision shield during execution. We demonstrate this approach boosts performance rather than solely using the learnt policy. We then show how we can more closely integrate learnt MAPF policies with LaCAM \cite{okumura2022lacam} (another heuristic search method) to further improve performance. We explore other variants of combining a learnt policy with PIBT and LaCAM and show significant improvement in success rate and costs over the learnt policy by itself.

Our goal in this paper is \textit{not} to show that ML approaches for MAPF are superior to classical heuristic search approaches.
Our objective instead is to show that given a learnt policy, we can use heuristic-search-based model-agnostic methods to significantly improve the performance of these policies. 
In regards to whether ML-based approaches or heuristic search approaches are better, we offer a more nuanced view.
We demonstrate in our experiments that given existing strong 2D heuristics (i.e. backward Dijkstra's), current classical heuristic search approaches are extremely strong and will likely outperform ML approaches in 2D MAPF.
However given imperfect heuristics, we see that ML approaches can actually outperform certain heuristic search methods. Section \ref{sec:learnt-policies-noise} discusses in further detail where learning may be applicable given our findings.

Overall, the main point we attempt to show in this paper is that ML methods for MAPF should fully leverage heuristic search. Doing so can substantially boost success rates and scalability as shown in Figure \ref{fig:boosting-success-rate}. Succinctly stated, our main contributions are:
\begin{enumerate}
    \item Creating a ``smart" collision shield using PIBT that effectively resolves learnt policy collisions instead of freezing colliding agents.
    \item Showing a neural network agnostic framework for using a learnt 1-step policy with PIBT/LaCAM for full horizon planning to enable theoretical completeness and boost success rate, and experimenting with several variants.
\end{enumerate}

\section{Related Works}
There are many different approaches for MAPF, ranging from optimization, heuristic search, to machine learning. We first define MAPF and then focus on the relevant heuristic search and machine learning approaches. 

\subsection{MAPF Problem Formulation}
We describe the classic single-shot 2D MAPF. Here, we are given a known gridworld with free space and obstacles and a single start-goal pair for each agent. Each agent can move in its 4 cardinal directions or wait for a total of 5 different actions per discretized timestep. Our objective is to find a set of collision-free paths for all the agents to reach their goals without obstacle collisions, vertex collisions (two agents at the same location at the same time), or edge collisions (two agents swapping locations across consecutive timesteps). In addition to finding a set of collision-free paths, we hope to find efficient paths which minimize the total flowtime (sum of each agent's path length until they rest at the goal).

Single-shot MAPF is harder than lifelong MAPF (where agents immediately move to a different goal if they reach their first goal) or disappear-at-goal MAPF variants as agents need to rest at the goal location. Agents waiting at the goal requires tough reasoning for learnt models as naively stopping these agents at the goal can block later agents from reaching their goal locations. Our framework is applicable to other MAPF variants too but we choose to evaluate it on the more difficult single-shot MAPF scenario.

\subsection{Heuristic Search Approaches}
Heuristic search methods aim to tackle the exponentially growing search space by intelligently leveraging the semi-independence of agents.
Conflict-Based Search is a popular state-of-the-art complete and optimal framework for MAPF \cite{sharon2015cbs}. This technique plans for each agent individually and then resolves collisions (conflicts in their terminology) iteratively by applying constraints and replanning. Improvements on this foundational technique have been shown to scale up to solving hundreds of agents optimally or bounded-suboptimally \cite{barer2014suboptimal,bpEli2015,li2021eecbs,ictsMDDSharon2013,rhcrLi2020,wdgLi2019}. These methods typically have good solution quality but computationally scale poorly as the number of agents increases.

\subsubsection{PIBT}
Recently, Priority Inheritance with Backtracking (PIBT) and its extensions \cite{pibt,okumura2022lacam} have shown how greedy heuristic search methods can scale extremely well at the expense of solution cost. 
Algorithm \ref{alg:alg-pibt} describes PIBT which takes in agents' current states and priorities and computes/reserves the next 1-step move for each agent sequentially (Line \ref{line:priorities}). Each agent greedily attempts its best action but is prevented from colliding with static obstacles or reserved states/edges from higher priority agents (Line \ref{lines:collisions}). If an agent moves into a state occupied by another agent (Line \ref{lines:lower-agent}), the other (lower priority) agent is required to plan (thus ``inherits" the higher agent's priority) (Line \ref{lines:recusive-call}). This procedure repeats until a collision-free set of actions is found, if not successful then the first agent is forced to attempt its second-best action and logic repeats accordingly (Line \ref{lines:all-actions}).
PIBT interleaves greedy 1-step planning and execution but is still effective in long-horizon MAPF tasks. PIBT uses a backward Dijkstra's heuristic to determine what actions are best (i.e. the action leading to the state with the least heuristic estimate is the best). A backward Dijkstra's computes the optimal cost for each state by running a Dijkstra's starting at the goal state.

\begin{algorithm}[t]
\caption{PIBT}
\label{alg:alg-pibt}
\textbf{Parameters}: Current states $s^{1:N}$, actions $a^{1:N}_{1:5}$, Static Obstacles $W$ \\ \noindent
\textbf{Output}: Collision free actions $a^{1:N}$
\begin{algorithmic}[1] 

\Procedure{PIBT}{Current states $s^{1:N}$, agentPriorities}
    \State Reserved = $\emptyset$
    \State Moves = dictionary()
    \For{agent $k \in $ argsort(agentPriorities)} \label{line:priorities}
        \If {$k \notin $ Moves.keys()}
            \State PIBT-H($k$, $a^{1:N}_{1:5}$, $s^{1:N}$, Reserved, Moves)
        \EndIf
    \EndFor
    \State \textbf{return} Moves
\EndProcedure
\Statex

\Procedure{PIBT-H}{Agent $k$, Actions $a^{1:N}_{1:5}$, Current States $s^{1:N}$, Reserved $RS$, Moves}
    \For{$a \in a^k_{1:5}$} \Comment{Sort $a^k_{1:5}$ with preferred $a$ first} \label{lines:all-actions}
        \State $s' \gets T(s^k,a)$
        \If  {$s' \in W \cup RS$ or $(s',s) \in RS$} \label{lines:collisions}
            \State Continue
        \EndIf
        \State Moves[$k$]$=a$, $RS \pluseqB \{s,(s,s')\}$)
        \If {$\exists$ agent $j \neq k$ at $s'$} \label{lines:lower-agent}
            \If {PIBT-H($j$, $a^{1:N}_{1:5}$, $s^{1:N}$, $RS$, Moves)} \label{lines:recusive-call}
                \State \textbf{return} Success
            \EndIf
            \State Moves[$k$]$=\emptyset$, $RS \minuseqB \{s, (s,s')\}$)
        \Else
            \State \textbf{return} Success
        \EndIf
    \EndFor
    \State \textbf{return} Failure
\EndProcedure
\end{algorithmic}
\end{algorithm}

\subsubsection{LaCAM} \label{sec:lacam}
Lazy Constraints Addition search for MAPF (LaCAM) \cite{okumura2022lacam} builds on PIBT by using it as a successor generator within a Depth-First Search (DFS) of the joint-configuration space. We describe LaCAM in a simplified way to get the main idea across, but note that LaCAM is more nuanced. Given $N$ agents, imagine running a ``joint-configuration" space DFS. Specifically, given the initial configuration, we generate all possible valid neighboring joint-configuration successors, pick one we have not seen, and repeat. Note that a valid joint-configuration successor will move agents by only one step (or wait) and will not have vertex, edge, or obstacle collisions. In 2D MAPF with each agent having $5$ actions, this means $N$ agents can generate up to $5^N-1$ new successors (the minus 1 as all agents waiting will not result in a new configuration). This approach is clearly not scalable to many agents due to the exponential number of successors.

LaCAM mitigates this issue by generating successors lazily. The key insight is that given a joint-configuration $J_A$, we must generate all possible successors, but we can do this sequentially rather than all at once. They do this by employing lazy constraints where each constraint specifies that an agent should be at a specific location. If we encounter $J_A$ multiple times when back-tracking in the DFS, we require different agents to be at different locations and generate different joint-configuration successors satisfying these constraints. Figure \ref{fig:main-figure} shows an example where DFS goes to the left, exhausts successors, and then revisits the start configuration and lazily generates a new successor to the right (e.g. by arbitrarily choosing to constrain the orange agent to be in the middle). Given sufficient time and memory, LaCAM explores all possible successors of $J_A$. 
It is therefore crucial that the configuration generator be fast while satisfying the constraints. 
LaCAM found that PIBT performed well as a configuration generator compared to other MAPF methods. Since LaCAM eventually explores all configurations, LaCAM is theoretically complete. 
In practice, LaCAM improves success rate over PIBT as the DFS over configurations allows getting out of local minima (e.g. deadlock). 
LaCAM is extremely effective on existing 2D benchmarks.


\subsection{Machine Learning Approaches} \label{sec:ml-related-work}
Machine learning approaches typically attempt to learn local 1-step policies for each agent which they then execute in parallel to solve the MAPF instance. Formally, a local policy $\pi$ for agent $i$ learns a conditional action probability distribution $p_\pi(a^i_{1:5}|\mathcal{O}^i)$. $\mathcal{O}^i$ is a local observation that typically contains local (e.g. 3x3 or 11x11) obstacle and goal information around agent $i$'s state, as well as information about nearby agents (including possibly inter-agent communication).
Given the action distributions, agents typically take the highest probable actions (or sample instead) and execute them as long as they are collision-free. Since the learnt network is not perfect, some proposed actions may not be collision-free and need to be resolved for feasibility.


PRIMAL \cite{sartoretti2019primal} is a foundational machine learning method that uses reinforcement learning and supervised learning to learn local policies. PRIMAL2 \cite{damani2021primal2} improves the observation inputs from PRIMAL to handle mazes by automatically annotating potential bottlenecks. SCRIMP \cite{wang2023scrimp} uses the same framework but replaces the complicated PRIMAL2 inputs with extremely small 3x3 observations and inter-agent communication via a modified transformer.

GNN \cite{li2020gnn} is a popular approach that solely uses supervised learning to learn a local policy. They use a Graph Neural Network \cite{graph_nn_2017} for communication and symmetry breaking across agents. MAGAT \cite{li2021magat} improves upon the message-passing neural network architecture in GNN. GNN introduces (and MAGAT also uses) a ``collision shield" which takes each agent's preferred action and executes it if collision-free or freezes the agent if it would cause a collision. Other works, e.g. PRIMAL and PRIMAL2, employ similar concepts by removing agents' actions that collide with other agents.

A key problem highlighted by most ML methods is deadlock between agents. PRIMAL's results show how deadlock can commonly occur when agents rest at their goal locations. PRIMAL2 specifically aims to learn conventions to decrease deadlock in maze structures. SCRIMP uses their learnt state values (an additional output of their model apart from their local policy) to calculate agent priorities that they use to prioritize agents in deadlock and show that this improves performance over naive collision shielding. We qualitatively noticed how MAGAT encounters deadlock as well. 
Our insight is that we can resolve local deadlock using heuristic search methods on top of the learnt policy predictions.


\section{Improving Learnt Local Policies with Heuristic Search}
One main motivation for learning MAPF policies is that each agent can run in a decentralized manner.
Theoretically, these techniques should be able to scale well to an increasing number of agents as the neural network inference time should be roughly constant regardless of the number of agents. However, existing ML methods execute 1-step policies that get stuck in deadlock/live-lock because they lack full horizon planning and do not have theoretical completeness guarantees. Thus, in practice, they have poor success rates and scalability in scenarios with larger agent congestion.

\begin{figure*}[t]
    \centering
    \includegraphics[width=0.97\textwidth]{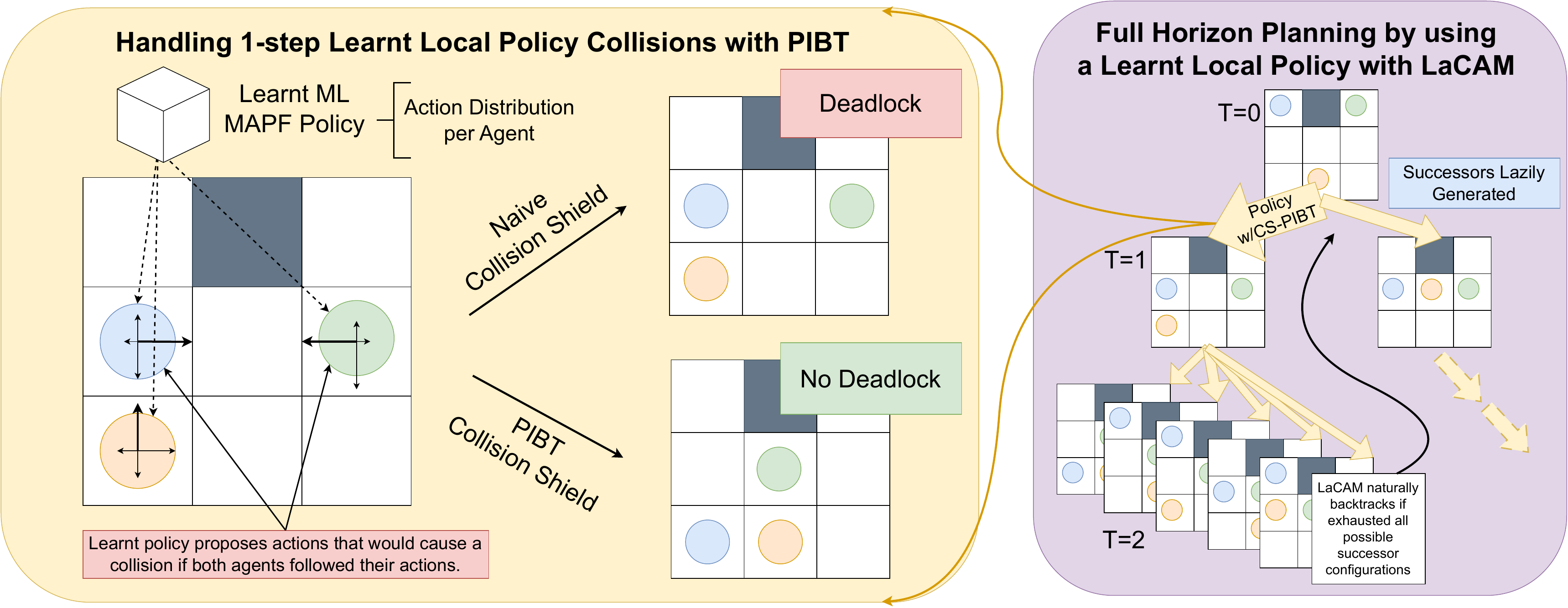}
    \caption{Given a learnt local MAPF policy which returns 1-step action distributions (depicted as black arrows overlaid on colored agents with larger magnitude depicting higher probability), we need to resolve collisions that might occur if we followed the proposed actions. We depict an example where blue and green would collide with each other. Existing work uses a ``naive collision shield" which only uses the agents' picked actions and replaces collisions with wait actions, which can cause deadlock between agents. We propose using the PIBT collision shield (CS-PIBT) to resolve 1-step collisions and reduce deadlock. Note that CS-PIBT uses the entire action distribution of the agent. To enable full horizon planning, we can use the LaCAM framework with the learnt policy with CS-PIBT as the configuration generator as defined in \citet{okumura2022lacam}. LaCAM in essence conducts a DFS over the joint-configuration space, enabling it to escape local minima by backtracking and improving success rates.}
    \label{fig:main-figure}
    \vspace*{-1em}
\end{figure*}

\subsection{Handling 1-step Collisions with PIBT} \label{sec:cs-pibt}
One fundamental issue with using learnt 1-step policies is that agents' learnt policies can choose actions that lead to collisions. Figure \ref{fig:main-figure} shows how two agents could choose to move into the same empty cell and lead to a collision if unchecked. Existing 1-step MAPF ML methods avoid this by post-processing actions by freezing agents that propose actions that collide. GNN \cite{li2020gnn} introduces this process as ``collision shielding".

\subsubsection{CS-NAIVE}
Formally, collision shielding is a function that takes in the current states and proposed action distribution and returns the next valid states that avoids vertex, edge, and obstacle collisions. \citet{li2020gnn} describe the following collision shielding process we term ``CS-Naive". Given $N$ agents in at states $s^{1:N}$ and actions $a^{1:N}$, we simulate $(s^i,a^i) \rightarrow s_{new}^i$ and detect collisions. All agents with collisions are told to wait at their current state.
Note that we must update $a^{1:N}$ accordingly and repeat this procedure as the added waits could cause additional conflicts (e.g. orange agent in Figure \ref{fig:main-figure} would collide with blue), and this process could iterate several times. 
As the authors observe, this can cause deadlock if multiple agents repeatedly propose actions that lead to collisions.
Figure \ref{fig:main-figure} shows an instance where collision shielding is necessary, and how two agents colliding can cause other agents to be stuck in deadlock.

\subsubsection{CS-PIBT}
One critical observation is that this collision checking does not take the full probability distribution of the agents' proposed actions. Specifically, we either take the chosen action or we wait, we never consider the other actions. Suppose we have two agents at different locations proposing to go to the same location, what if we could let one go there and the other agent pick its second best action? This could significantly reduce idling and deadlock. But now the question arises of which agent can attempt its primary action and which agent needs to try its second action. Priority-based techniques naturally solve this question by assigning priority to agents and having the better priority agent have precedence. We thus want a single-step priority-based approach that takes in a preference of actions and returns a valid configuration. PIBT does exactly this.

Concretely, instead of solely taking in the single proposed action per agent, we propose feeding the entire probability distribution into PIBT along with agent priorities. PIBT naturally resolves obstacle, vertex, and edge collisions using agent priorities, backtracking, and the full action set. We term this use of PIBT as a collision shield for a learnt policy as ``CS-PIBT". CS-PIBT will always return a valid configuration as all agents waiting is a valid collision-free option. PIBT can be decentralized \cite{pibt}, and CS-PIBT is decentralized similarly to CS-Naive in that only colliding agents need to coordinate. We maintain the same iterative 1-step planning and execution process as before.

One small detail is that PIBT does not take in an action distribution but rather a strict action ordering (e.g. which action to try 1st, 2nd, ... 5th). We can convert an action probability distribution to an ordering by preferring actions in order of highest probability. Interestingly, as discussed in Section \ref{sec:exp-cs-pibt}, we found that although this improves performance over CS-Naive, this strict ordering is a large bottleneck. We can get much larger performance benefits by converting our action distribution into an action ordering by sampling from the probability distribution without replacement. 
Algorithm \ref{alg:alg-cs-pibt} describes CS-PIBT with the two possible variants.

\begin{algorithm}[t]
\caption{Collision Shield PIBT (``CS-PIBT")}
\label{alg:alg-cs-pibt}
\textbf{Parameters}: Current states $s^{1:N}$, action probability distribution $p^{1:N}_{1:5}$. Note $\|p^i\|$ is 5 as in 2D MAPF we have 4 movements along with waiting. $p^i_{1:5}$ can be from an arbitrary learnt model. \\ \noindent
\textbf{Output}: Collision free actions and states
\begin{algorithmic}[1] 
\Procedure{Get\textbf{Strict}ActionOrdering}{$p^i_{1:5}$}
    \State \textbf{return} action ordering sorted by decreasing $p^i_{1:5}$
\EndProcedure

\Procedure{Get\textbf{Sampled}ActionOrdering}{$p^i_{1:5}$}
    \State \textbf{return} reorder action orders by $p^i_{1:5}$ biased sampling without replacement
\EndProcedure

\Procedure{CS-PIBT}{$s^{1:N}, p^{1:N}_{1:5}$}
    \State $a^i_{1:5} \leftarrow$ Get\textbf{Sampled}ActionOrdering($p^i_{1:5}$) $~\forall i$
    \State $(s^{1:N}_{new}, a^{1:N}_{best}) \leftarrow $ PIBT($s^{1:N}, a^{1:N}_{1:5}$)
    \State \textbf{return} $(s^{1:N}_{new}, a^{1:N}_{best})$
\EndProcedure
\end{algorithmic}
\end{algorithm}


\subsection{Full Horizon Planning with LaCAM}
The previous subsection describes a PIBT-based ``collision shield" that handles 1-step collisions. However, using a learnt 1-step policy with this collision shield does not enable any theoretical guarantees as the 1-step policy can still get stuck in deadlock or be arbitrarily bad. Ideally, we would like to use a learnt 1-step policy in a manner that maintains theoretical solution guarantees.


Section \ref{sec:lacam} describes in depth how normal LaCAM works. The key observation is that LaCAM requires a fast configuration generator that can satisfy lazily added constraints, and that PIBT satisfies this. We observe that a learnt MAPF policy with CS-PIBT can be modified to be a valid configuration generator. Specifically, in order to work with LaCAM, our learnt model with collision shielding needs to handle constraints that get lazily added. Given constraints on agents, we can easily do this by invalidating proposed actions that violate constraints and having CS-PIBT only consider the valid subset (the same way as is done in PIBT within regular LaCAM). Another perspective is that instead of LaCAM using PIBT informed by a heuristic, we can use PIBT informed by the learnt policy action distribution.

Since the learnt policy in this approach only reorders action preferences, it only alters the order in which configurations are searched and does not prune out any configurations. Thus using a 1-step learnt policy with CS-PIBT within LaCAM enables full horizon planning with the same completeness guarantee as LaCAM, because the joint-algorithm will still exhaustively search all configurations eventually. Using LaCAM does require a centralized search effort wrapping over the decentralized model with CS-PIBT, but this centralized structure is significantly weaker than other MAPF methods like EECBS \cite{li2021eecbs} or MAPF-LNS2 \cite{li2022mapf-lns2} which requires iterations of sequential replanning of agents. 
We note there currently exists no decentralized MAPF method that ensures completeness.

\subsection{Combining a Local Policy with a Heuristic} \label{sec:combining}
As mentioned, one perspective of using a learnt model with CS-PIBT as a configuration generator in LaCAM is that we are using regular PIBT informed by the learnt local policy $\pi$ instead of by a cost-to-goal heuristic $h(s)$. This perspective implies that some middle ground is possible; we can use PIBT informed by both the heuristic as well as the learnt local policy. This can be done in various manners.

Concretely, in regular PIBT, an agent at $s$ picks the action $\min_a h(s'=T(s,a))$ where $s'$ is a successor state of $s$ after applying the transition function $T(s,a)$. In 2D MAPF, there are many instances where two actions are tied and lead to different $s'$ which have the same minimum value, and \citet{pibt} show that tie-breaking randomly is important for good performance. On the other hand, our local policy with PIBT collision shielding picks the action $\max_a p_{\pi}(a|\mathcal{O})$. 
Note for simplicity we discuss how PIBT ``picks" its best action, but as mentioned earlier PIBT tries out actions in the corresponding order and picks the best action that is collision free.

Our first observation is that, instead of breaking ties randomly in PIBT between two equally good $h(s')$ states, we can break ties using $p_\pi(a|\mathcal{O}))$. 
We thus pick the lexicographic best action $\min_a (h(T(s,a)), 1-p_\pi(a|\mathcal{O}))$.
However, this tie-breaking mechanism is not satisfactory as it only uses the learnt policy sparingly in ties. We can generalize the utility of the learnt policy by picking $\min_a h(T(s,a)) + R \times (1 - p_\pi(a|\mathcal{O}))$ where $R$ is a hyper-parameter that weighs on how much we want to follow our policy over the heuristic.

Altogether we have four possible ways of combining the local policy with a heuristic:
\begin{align}
&    O_{h} = \min_a h(T(s,a)) \\
&    O_{\pi} = \max_a p_\pi(a|\mathcal{O}) \\
&    O_{tie} = \mathrm{lexmin}_a (h(T(s,a)), 1-p_\pi(a|\mathcal{O})) \\
&    O_{sum}(R) = \min_a h(T(s,a)) + R * (1 - p_\pi(a|\mathcal{O}))
\end{align}
Note, $O_{sum}(R=0)$ equals $O_{h}$ and $O_{sum}(R \rightarrow \infty)$ is identical to $O_{\pi}$.
In 2D MAPF with unit actions and a backward Dijkstra heuristic, given a state $s$, neighboring states $s'$ have heuristic values $h_{BD}(s') \in \{h_{BD}(s)-1, h_{BD}(s), h_{BD}(s)+1\}$. By definition, $p_\pi(a|\mathcal{O}) \in [0,1]$. Thus $O_{sum}(R)$ with $R \in (0,1]$ is equivalent to $O_{tie}$. 


\section{Experimental Results}
We have described multiple techniques to boost the performance of a learnt local MAPF policy with heuristic search. We seek to evaluate the following experimentally:

\begin{enumerate}
    \item How does CS-PIBT compare against naive collision shielding?
    \item How does integrating a learnt policy with LaCAM boost performance?
    \item What is the best way of combining a learnt policy with heuristic information?
\end{enumerate}

\begin{table*}[t!]
\centering
\resizebox{0.99\textwidth}{!}{

\begin{tabular}{|c|c|c|c|c||c|c|c|c|c|c|c|c|c|}
\multicolumn{14}{c}{Success Rate} \\ \hline
 & \multicolumn{2}{c|}{$h_{Manhattan}$} & \multicolumn{2}{c|}{$h_{BD}$} & \multicolumn{5}{c|}{MAGAT with} & \multicolumn{4}{c|}{Simple Policy $\pi_s$ with} \\ \hline
Agents & PIBT & LaCAM & PIBT & LaCAM & CS-Naive & \multicolumn{1}{c}{CS-PIBT} & w/$O_{tie}$ & \multicolumn{1}{c}{LaCAM} & w/$O_{tie}$ & CS-Naive & CS-PIBT & \multicolumn{1}{c}{LaCAM} & w/$O_{tie}$ \\ \hline
50 & 0 & 1 & 0.98 & 1 & 0.90 & 1 & 1 & 1 & 1 & 0 & 0.92 & 1 & 1 \\
100 & - & 1 & 0.98 & 1 & 0.42 & 0.97 & 0.92 & 1 & 1 & - & 0.88 & 1 & 1 \\
200 & - & 1 & 0.83 & 1 & 0 & 0.59 & 0.41 & 0.87 & 0.8 & - & 0.59 & 0.99 & 0.99 \\
300 & - & 1 & 0.55 & 1 & - & 0 & 0.28 & 0 & 0.72 & - & 0 & 0.89 & 0.92 \\
400 & - & 0.95 & 0.4 & 1 & - & - & 0 & - & 0.6 & - & - & 0.57 & 0.68 \\ \hline
\multicolumn{14}{c}{} \\ 
\multicolumn{14}{c}{Average Path Cost (Length) per Agent over Successful Instances} \\ \hline
50 & - & 132 & 25.9 & 25.7 & 25.6 & 23.5 & 22.7 & 23.7 & 22.7 & - & 26.2 & 26.3 & 25.6 \\
100 & - & 232 & 28.4 & 28.7 & - & 27.2 & 25.0 & 27.0 & 25.0 & - & 29.6 & 29.1 & 28.5 \\
200 & - & 402 & 34.5 & 34.7 & - & 41.1 & - & 39.4 & 30.1 & - & 34.7 & 34.6 & 33.7 \\
300 & - & 617 & 40.5 & 40.8 & - & - & - & - & 35.4 & - & - & 40.7 & 39.4 \\
400 & - & 907 & 49.5 & 49.3 & - & - & - & - & 45.2 & - & - & 50.0 & 47.4 \\ \hline
\end{tabular}}
\caption{The top table compares the success rates of using different learnt local policies with our proposed methods. We include PIBT and LaCAM runs using a Manhattan ($h_{Manhattan}$) and backward Dijkstra's ($h_{BD}$) heuristic as baselines. Existing works using a naive collision shield (CS-Naive) to handle 1-step collisions, we propose using a PIBT-based collision shield (CS-PIBT) and incorporating the policy with LaCAM.
We finally evaluate how combining the learnt policy with $h_{BD}$ through $O_{tie}$ affects performance. We see that using CS-PIBT, LaCAM, and $O_{tie}$ significantly boosts performance without any changes to the model.
The bottom table shows the average path cost per agent across instances that all methods with success $\geq 50\%$ were able to solve.
Note MAGAT $O_{tie}$ is MAGAT with CS-PIBT with actions ordered by $O_{tie}$, while the corresponding column for $\pi_s$ uses LaCAM with $O_{tie}$. We see that combining both together leads to better costs than using just the policies themselves.
}
\label{tab:mainResults}
\vspace*{-1em}
\end{table*}


\subsection{Learnt Policies used for Evaluation}
All of our techniques are agnostic to the methodology or architecture of the learnt policy.
We implemented and incorporated PIBT collision shielding and LaCAM for the state-of-the-art pre-trained MAGAT neural network. Note to practitioners: Implementing CS-PIBT is easy given that it is a post-processing technique, but implementing LaCAM requires more work and interfacing between the model and search.
We additionally trained a significantly simpler network ourselves to evaluate the impact of our methods on a weaker model. 
We test on the standard random-32-32-10 map \cite{stern2019mapfbenchmark} as MAGAT is trained on environments with 10\% randomly sampled obstacles. All results are aggregated across 25 provided scenarios and 5 seeds, and all methods attempt to minimize the sum of agents' path lengths. For CS-PIBT and LaCAM, we used dynamic priorities as defined in \citet{pibt} which assigns initial high priorities to agents far from their goals and then sets priorities to zero when agents reach their goals.

\subsubsection{MAGAT}
MAGAT is a state-of-the-art model that uses a graph neural network structure and supervised learning for one-shot MAPF \cite{li2021magat}. MAGAT was shown to scale well given similar agent densities, i.e. they show that MAGAT trained on 20x20 maps with 10 agents (density of 0.025) is able to scale to 200x200 maps with 1000 agents (same 0.025 density) with an 80\% success rate. Note that most classical MAPF methods evaluate performance by increasing the number of agents in the same map, i.e. increasing agent density. Given a 50x50 map, MAGAT evaluates performance up to 100 agents (density 0.04) and starts to show performance degradation after 60 agents (density 0.024). They do not show results past 0.04 density, implying MAGAT fails past this level. For context, existing heuristic search methods like EECBS, PIBT, and LaCAM can work in 40+\% agent densities.

\subsubsection{Simple Learnt Policy} 
The objective of this simple learnt policy is to see how our techniques can improve relatively weak models. 
We intentionally train a simple policy to contrast MAGAT and avoid extra hyper-parameters and hyper-parameter tuning.

We trained our simple policy $\pi_s$ similar using supervised learning. Concretely, we ran EECBS on the random-32-32-10 map for 20-200 agents and collected each state-action pair as a training example as an educated full horizon oracle for our model. For each agent $i$ with (state, action), we fed in 6 local 9x9 fields of view centered at the state. The first 9x9 was the local map (0's free space, 1 obstacle) and the 2nd 9x9 was a local backward Dijkstra's heatmap (similar to PRIMAL2) guiding agent $i$ to its goal. The other four 9x9 channels were the backward Dijkstra's heatmap of the four closest agents within the field of view, with all-zero heatmaps if not enough agents. This is fairly in line with existing work. 

The main changes in respect to MAGAT and similar methods are that our neural network is significantly smaller than all existing work (just 1 CNN layer, 2 MLP layers) and critically has no inter-agent graph neural network or communication structure. After the CNN encoding, we also add in the coordinates of the four closest agents relative to agent $i$'s current location (or pad with zeros accordingly). Due to its small nature, the model we trained was unable to generalize to new maps so we just trained on the same map we tested on with a subset of agents and start-goals. 
As seen later on, the simple policy cannot succeed past 10 agents with CS-Naive. However, with CS-PIBT and our LaCAM framework, our policy can scale up to 200 and 400 agents respectively. This implicitly reveals how integrating heuristic search can potentially allow ML practitioners to learn simpler models (with potential computational benefits) rather than existing larger models.

\subsection{Improving Scalability and Success Rate}
\subsubsection{CS-PIBT} \label{sec:exp-cs-pibt}
Table \ref{tab:mainResults} evaluates the effectiveness of the collision shield, comparing CS-Naive to CS-PIBT on MAGAT and our simple policy. All the methods were given a 60-second timeout except for MAGAT which used the maximum makespan limit described in their paper. MAGAT with CS-Naive has good success at 50 agents ($\approx$ agent density 0.05) consistent with MAGAT's results, but struggles to scale larger. Using the same model predictions with CS-PIBT, we see that it is able to have decent success rate on 200 agents. This is a large improvement in agent density scalability with the exact same model.
Likewise, our simple policy is unable to solve 50 agents, and upon inspection can only solve up to 10 agents well. With CS-PIBT, it can scale up to 200 agents (over a 10x improvement in scalability). 
These results convincingly demonstrate that CS-PIBT can significantly improve success rates and scalability in congestion. When visualizing the different collision shields, we noticed that CS-PIBT has two main strengths over CS-Naive. First, CS-PIBT allows robust movement when several agents are grouped up in the same area, while in CS-Naive mainly agents on the borders moved while the internal ones took longer to get out. Second, CS-PIBT allows agents to more easily go through agents resting on their goal, which was common in most failures in CS-Naive for a low number of agents and CS-PIBT at higher numbers. Both these behaviors are direct results of CS-PIBT's usage of priority inheritance and taking the full action probability distribution into account.
A small but crucial detail for using CS-PIBT as discussed in the next section is that the action probability should employ randomness.

One note on runtime is that the overhead of CS-PIBT is fairly negligible based on implementation. For $\pi_s$, CS-PIBT was implemented in C++ and took $0.05$ milliseconds per CS-PIBT call for 200 agents and had a negligible impact on overall runtime which was dominated by neural network input and inference time. CS-PIBT in Python for MAGAT took about $3$ milliseconds per call for 200 agents which roughly translates to 3\% of total runtime.

\begin{figure}[t]
    \centering
    \includegraphics[width=0.45\textwidth]{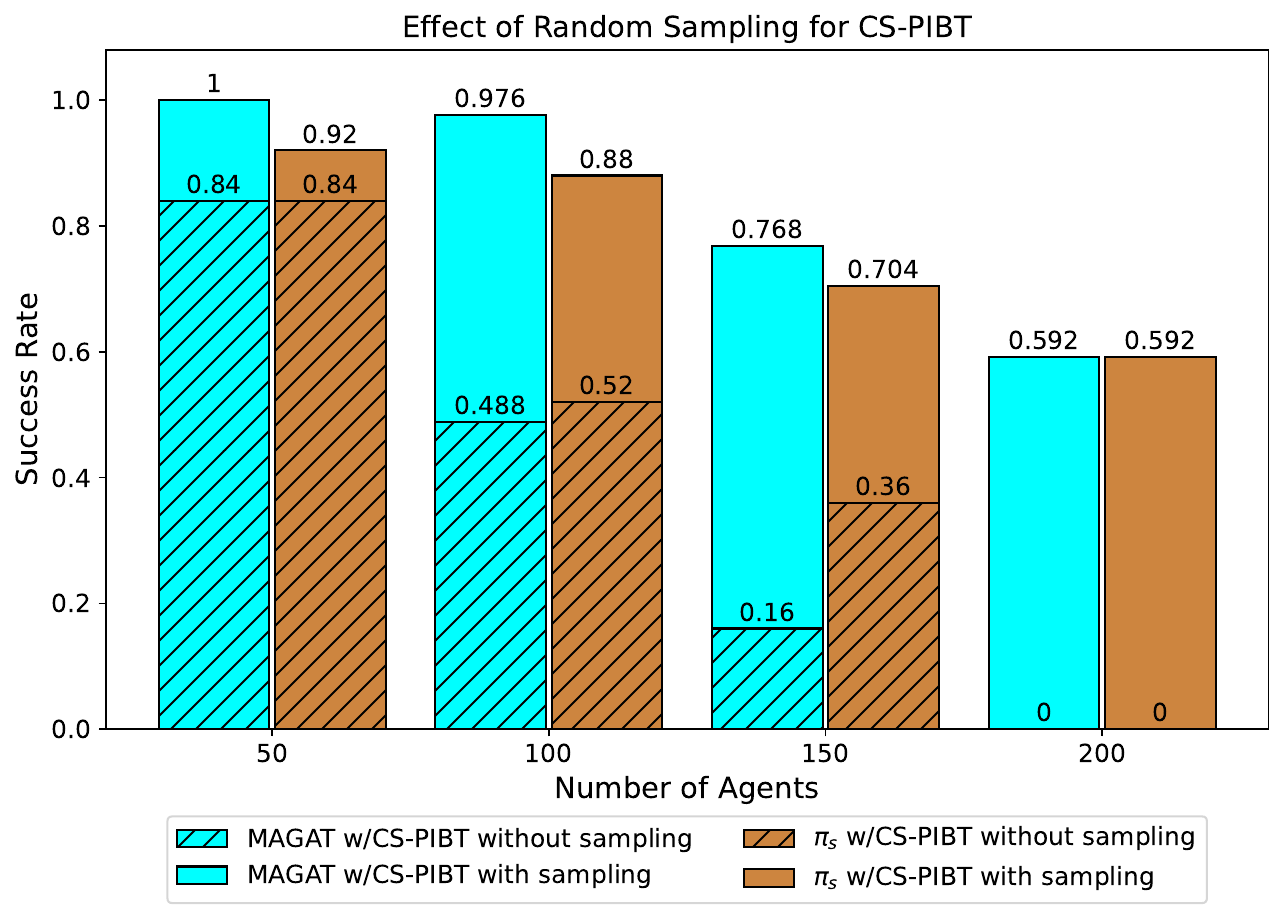}
    \caption{We plot the effect of using CS-PIBT with and without biased sampling. We see that including sampling significantly improves performance instead of always choosing actions with the highest probability first. 
    }
    \label{fig:random-sample}
    \vspace*{-1.5em}
\end{figure}

\begin{figure}[t]
    \centering
    \includegraphics[width=0.45\textwidth]{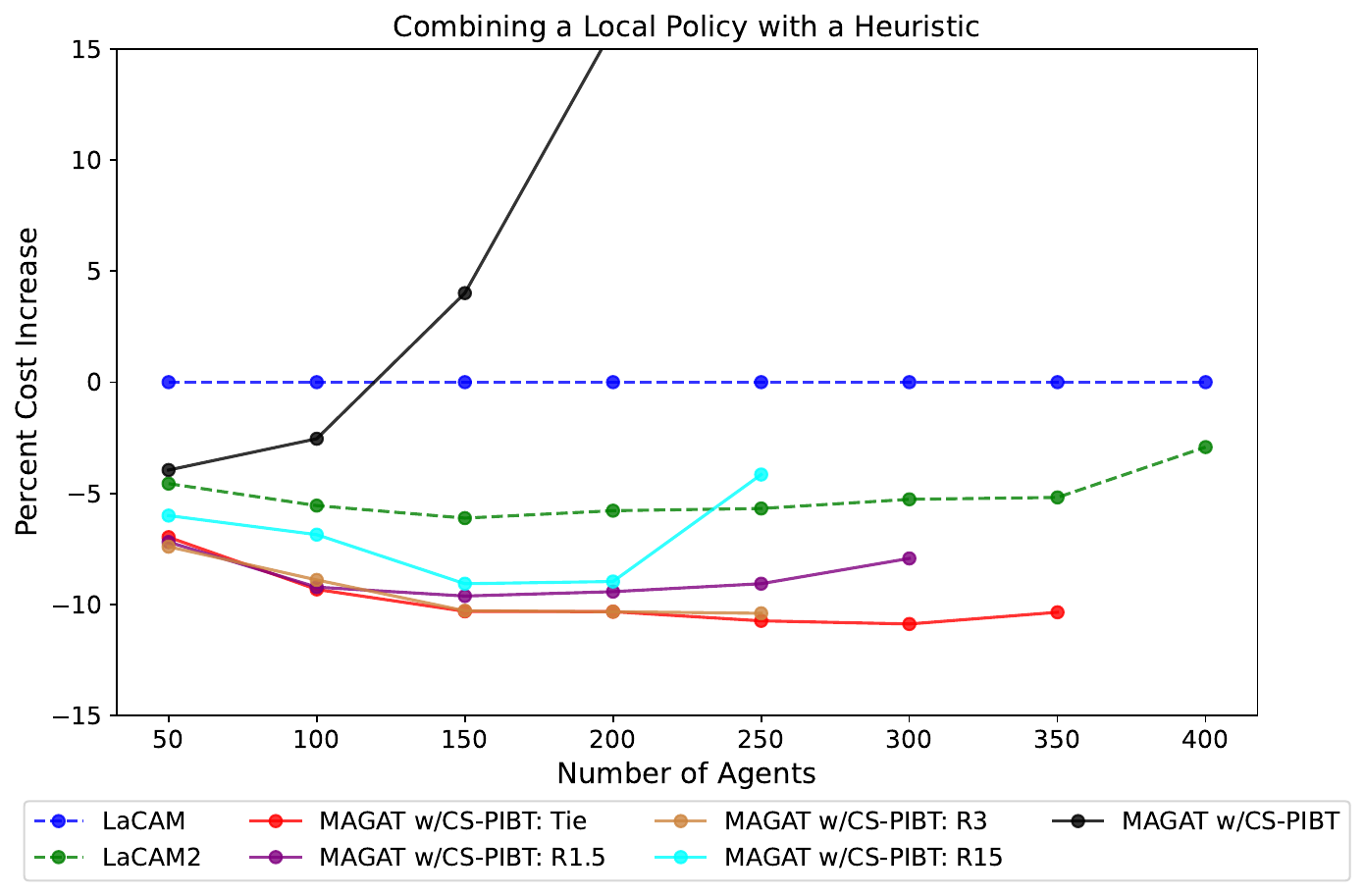}
    \caption{$\downarrow$ is better. We evaluate different methods of combining MAGAT's local policy with the standard backward Dijkstra's heuristic used in LaCAM. We evaluate the cost differences with respect to regular LaCAM informed by $h_{BD}$ and random tie-breaking (blue). LaCAM2 (green) tie-breaks preferring locations without other agents which was shown to improve solution cost (but reduce success rate) in \citet{pibt}. ``Tie" (red) tie-breaks $h_{BD}$ by using MAGAT's preferences. MAGAT (black) disregards $h_{BD}$. Intermediate ``R" methods (purple, orange, cyan) sort using a weighted combination of $h_{BD}$ and MAGAT's probabilities. We see that tie-breaking improves solution cost over LaCAM, LaCAM2, and MAGAT.
    }
    \label{fig:combining}
    \vspace{-1em}
\end{figure}

\subsubsection{The Importance of Randomness}
The authors of PIBT mention how random tie-breaking plays an important part in improving performance (which we also found when testing out baseline PIBT). 
Interestingly, we find that employing randomness for converting the policy's action distribution into an action ordering is crucial to our CS-PIBT. Figure \ref{fig:random-sample} shows the effect of using CS-PIBT using strict ordering vs sampled action ordering (see Algorithm \ref{alg:alg-cs-pibt}). We see that strict ordering is strictly worse than sampling and that its performance boost degrades fast as the number of agents increases. 

One hypothesis to explain this difference is that the model could be biased and certain actions' probabilities could consistently be higher than another, e.g. Pr(up)=0.55 $>$ Pr(down)=0.45. In this case, strict ordering will always try up before down, but with biased sampling we will try up before down only 55\% of the time, matching the intended distribution. However, we analyzed the action ordering distributions induced by strict action orderings and sampled action orderings and found they had very similar distributions. Qualitatively, we observed that failure instances for CS-PIBT without sampling had many live-locks where two agents alternated back and forth between the same few states, and got stuck around obstacles. We did not notice this as often with CS-PIBT with sampling, implying that sampling helps the agent overcome live-lock by trying out different actions instead of repeating the same ones.

\subsubsection{Full Horizon Planning with LaCAM}
Table \ref{tab:mainResults} shows the result of running MAGAT and the simple policy $\pi_s$ within the LaCAM framework. We see that LaCAM improves the success rate for both, and increases scalability for $\pi_s$. LaCAM's overall framework allows searching over multiple options which enables the search to overcome deadlock and boosts success rate. 
These results are consistent with LaCAM's original results improving PIBT's 1-step planning. 
We highlight how $\pi_s$ + LaCAM significantly improves scalability as it solves over 300 agents while $\pi_s$ with CS-PIBT struggles at 200 while $\pi_s$ by itself cannot solve 20 agents.

\begin{figure*}[t!]
    \begin{subfigure}{0.32\textwidth}
         \centering
         \includegraphics[width=\textwidth]{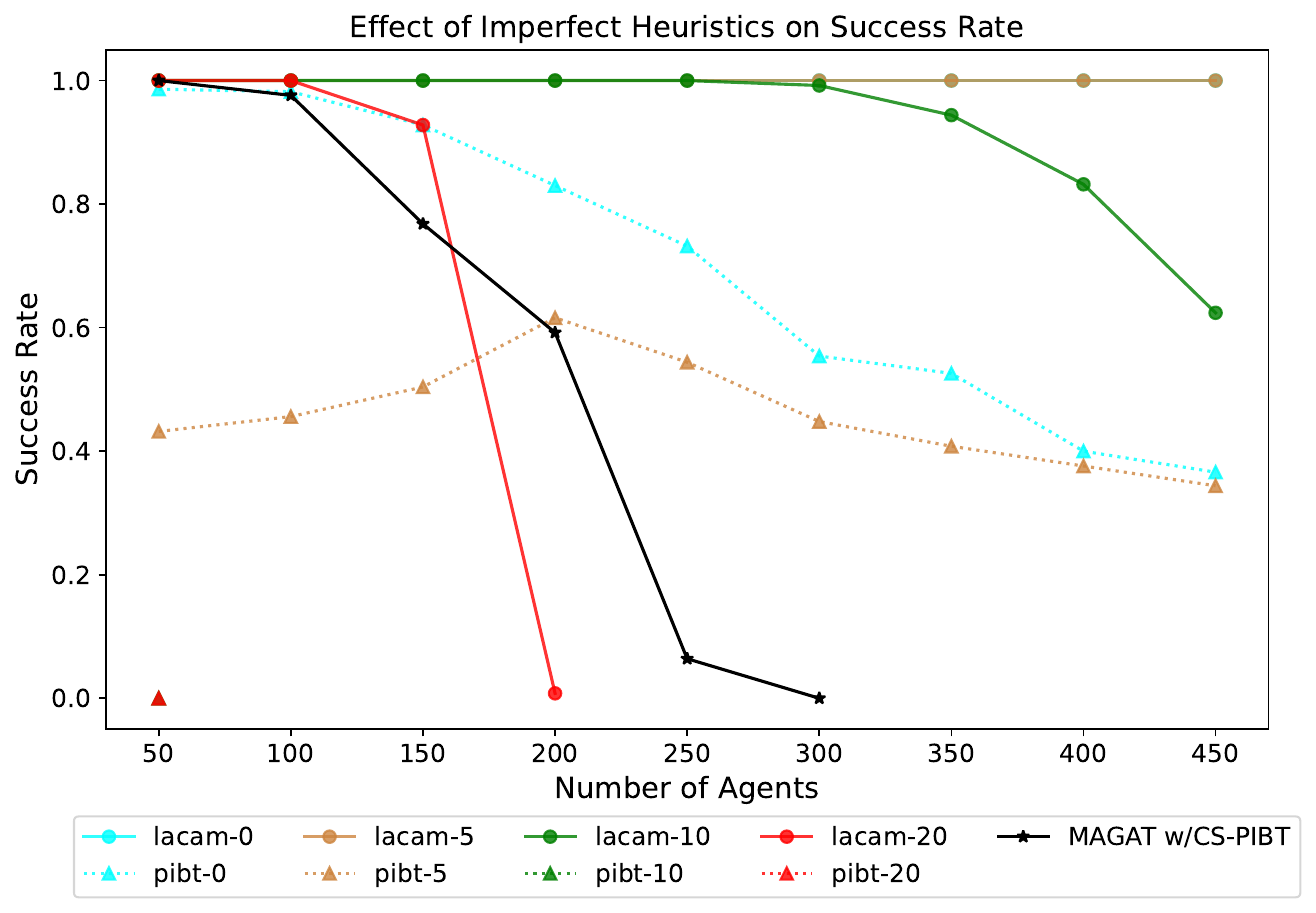}
         \subcaption{Success Rate}
         \label{fig:noise-success}
    \end{subfigure}
    \hfill
    \begin{subfigure}{0.32\textwidth}
         \centering
         \includegraphics[width=\textwidth]{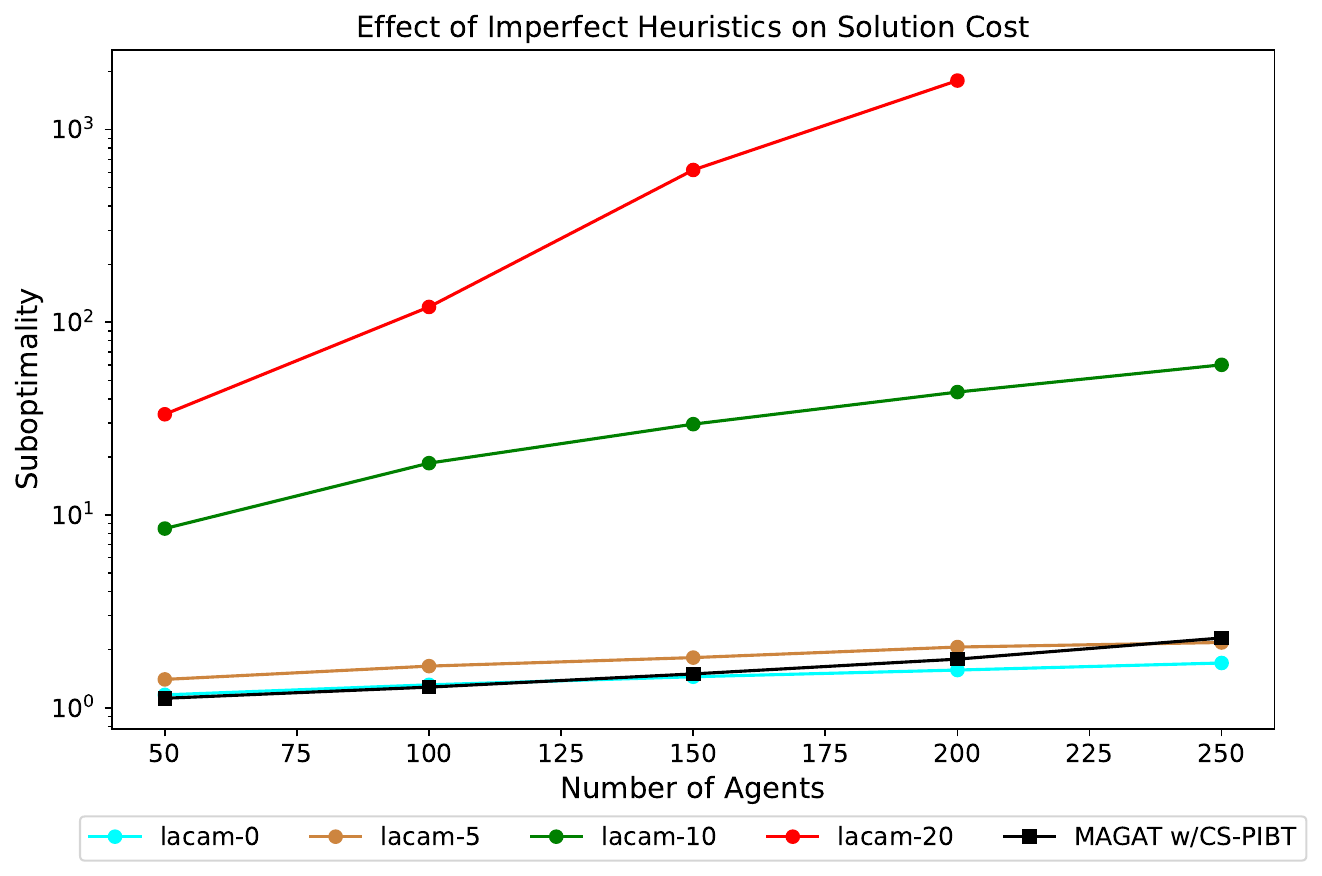}
         \subcaption{Solution Cost, 0-20\% $\bar{h}_{BD}$ imperfection}
         \label{fig:noise-cost-large}
    \end{subfigure}
    \begin{subfigure}{0.32\textwidth}
         \centering
         \includegraphics[width=\textwidth]{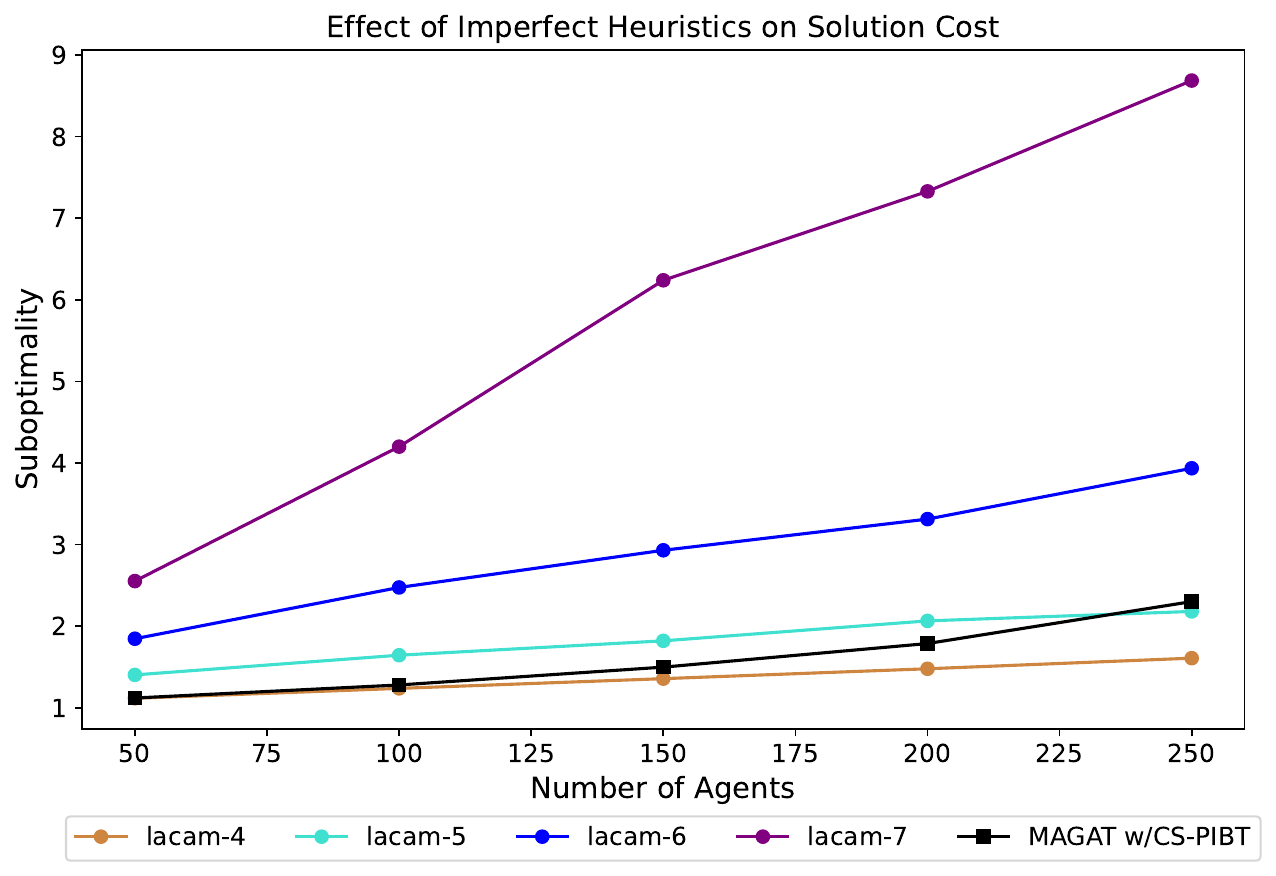}
         \subcaption{Solution Cost, 4-7\% $\bar{h}_{BD}$ imperfection}
         \label{fig:noise-cost-small}
    \end{subfigure}
    \caption{We explore the effect of imperfect heuristics on PIBT and LaCAM. Given a setting of ``$K\%$" imperfection and the perfect backward Dijkstra's heuristic $h_{BD}(s)$, we uniformly sample from $e \sim [1-K/100,1]$ and obtain our imperfect $\bar{h}_{BD}(s) = e \times h_{BD}(s)$ which adds noise to the calculated heuristic values to simulate uncertainty. (a) shows the performance with PIBT and LaCAM in $K=0$ (cyan), $5$ (brown), $10$ (green), and $20$ (red) $\bar{h}_{BD}$-imperfect settings. We additionally plot MAGAT which does not use $h_{BD}$ and is thus independent of the heuristic imperfections. 
    From the success rate, we see that PIBT completely fails starting at $K=10$ (hidden by red triangle) and LaCAM fails at $K=20$. However, the solution cost in (b) reveals a worse picture; even though LaCAM succeeds with $K=10$ or $20$, the solutions are extremely suboptimal with LaCAM finding 100 times worse solutions. (c) highlights that LaCAM can be extremely brittle with it performing reasonably at $6\%$ but producing substantially worse solutions at $7\%$.
    }
    \label{fig:noise-both}
    \vspace{-1em}
\end{figure*}

\subsection{Combining a Policy with a Heuristic} \label{sec:combining-results}
Figure \ref{fig:combining} explores the different methods of combining a policy with a heuristic. We compare LaCAM solution cost against MAGAT w/CS-PIBT with action preferences sorted as described in Section \ref{sec:combining}. Results are percent cost increases with respect to LaCAM on successful runs, thus negative values denote cost improvements. We included an additional LaCAM baseline (LaCAM2) which, when given two equally good actions, will tie-break to avoid agent collisions, which was shown in \citet{pibt} to improve cost at the expense of success rate. Note that LaCAM performs full horizon planning while all the MAGAT results are iteratively 1-step planning and executing using CS-PIBT.

We observe that using just MAGAT's policy (black) leads to worse solutions than LaCAM. However tie-breaking (red), which primarily uses $h_{BD}$ and tie-breaks using MAGAT's predictions, leads to a consistent cost improvement of around 10\%. Using $O_{sum}(R)$ with different $R$ usually interpolates between the two behaviors, with higher values of $R$ performing worse (brown $R=3$ is an outlier).
One possible hypothesis for $O_{tie}$'s better performance is that MAGAT's model simply prefers to avoid 1-step collisions. However, LaCAM2's tie-breaking mechanism does exactly this but $O_{tie}$ is still better than LaCAM2 by about 5\%, with the performance improvement increasing as the number of agents increases. This implies that MAGAT is learning something more nuanced than a simple rule to get performance benefits.

We observed similar trends with our simple policy $\pi_s$ with respect to using $O_{tie}$ and $O_{sum}(R)$, except since $\pi_s$ is weaker all costs are shifted up. $\pi_s$ with $O_{tie}$ still improves cost compared to LaCAM from 1\% to around 4\% with larger benefits as the number of agents increases, but it is worse than LaCAM2. $O_{sum}(R)$ with $\pi_s$ showed similar interpolation behavior as with MAGAT (except no $R=3$ outlier) and similarly leads to worse performance than $O_{tie}$.

\subsection{Should We Even Use Learnt Policies?} \label{sec:learnt-policies-noise}
One observation with Table \ref{tab:mainResults} is that LaCAM serves as an extremely strong baseline. LaCAM with the backward Dijkstra's heuristic ($h_{BD}$) has perfect success rate, good cost, and is extremely fast. LaCAM requires no complex training, hyper-parameters, and works on arbitrary 2D maps. All learnt MAPF works require non-trivial data collection and complex models but perform worse than LaCAM. 

Our solution quality results in Section \ref{sec:combining-results} show one possible benefit of using learnt models; we can use them with heuristic search to improve path costs. In scenarios where cost is important, these 5-10\% percentage improvements could be worth the complications of learning a policy.

But what if small solution cost differences are not worth the infrastructure required for learning a MAPF policy, and we primarily care about success rate? LaCAM seems clearly superior to existing learnt MAPF models in this case.

However, one crucial assumption in 2D MAPF heuristic search methods is that we have access to an extremely strong single-agent heuristic, the backward Dijkstra's heuristic. Given no inter-agent interactions, this heuristic is perfect. Figure \ref{fig:noise-both} shows that PIBT and LaCAM are extremely reliant on this heuristic. We compute a ``$K\%$" imperfect heuristic $\bar{h}_{BD}(s) = e \times h_{BD}(s)$ where $e \sim [1-K/100,1]$. Note this heuristic is still admissible. Figure \ref{fig:noise-success} shows that with 10\% imperfections, PIBT fails (bottom left covered by red triangle) and \ref{fig:noise-cost-large} shows LaCAM outputs extremely suboptimal to the extent of non-usable solutions. A finer manipulation into $K$ in Figure \ref{fig:noise-cost-small} shows that LaCAM is extremely brittle to heuristic imperfections as it succeeds reasonably with $K=6$ but produces highly suboptimal paths with $K=7$. Table \ref{tab:mainResults}'s PIBT and LaCAM results with $h_{Manhattan}$ show similar results when using a Manhattan distance heuristic. MAGAT is not reliant on $h_{BD}$ and outperforms PIBT and LaCAM for $K \geq 7$ or $h_{Manhattan}$.

Thus in 2D MAPF, given a perfect backward Dijkstra's heuristic, LaCAM's success rates are impressive and it seems very hard for a learnt local policy to beat it. 
However, in scenarios where we cannot obtain a perfect or nearly-perfect backward Dijkstra's heuristic, learnt policies have the potential to outperform heuristic search algorithms.
The particular 2D MAPF scenarios where we think learnt policies may be useful are instances where $h_{BD}$ cannot be computed, such as in partially observable MAPF (where only part of the map is observed), or in extreme lifelong MAPF where goals are frequently changing and the overhead for computing $h_{BD}$ becomes prohibitively expensive. 

Learnt policies instead will likely excel in high dimensional state-space problems required for more realistic robot models. For example, 2D warehouse agents are constrained to move in grids at unit velocity in standard 2D MAPF formulation, but in reality, can move at angles and non-unit velocities. Realistic MAPF would then need to solve at least a 4-dimensional problem (x,y,$\theta$,velocity). Computing a perfect backward Dijkstra's heuristic or other heuristics within 6\% imperfections will likely be impossible here. Learnt MAPF policies, combined with CS-PIBT or LaCAM, would be promising in these situations.

\section{Conclusion}
We showed several model-agnostic methods of improving learnt local MAPF policies with heuristic search. We first introduced CS-PIBT which is a collision shield using PIBT that takes in a policy's 1-step probability distribution and outputs valid collision-free steps. We demonstrated how this significantly improves scalability and success rate without changing the model. We then showed how we can use a learnt model with LaCAM to enable full horizon planning with theoretical completeness and in practice further boosts success and scalability. From our literature review, we have shown results for the first time where a MAPF learnt policy (with search) can scale to similar agent densities (20+\%) of classical heuristic search methods. 
We hope future methods that learn local MAPF policies utilize these methods, and that more broadly, researchers work more on improving learning with heuristic search techniques. \textbf{Acknowledgements} R.V. thanks A.L. for his motivation. This material is partially supported by NSF Grant IIS-2328671.


\bibliography{aaai24}

\begin{thebibliography}{17}
\providecommand{\natexlab}[1]{#1}

\bibitem[{Barer et~al.(2014)Barer, Sharon, Stern, and Felner}]{barer2014suboptimal}
Barer, M.; Sharon, G.; Stern, R.; and Felner, A. 2014.
\newblock Suboptimal variants of the conflict-based search algorithm for the multi-agent pathfinding problem.
\newblock In \emph{Seventh Annual Symposium on Combinatorial Search}.

\bibitem[{Boyrasky et~al.(2015)Boyrasky, Felner, Sharon, and Stern}]{bpEli2015}
Boyrasky, E.; Felner, A.; Sharon, G.; and Stern, R. 2015.
\newblock Don’t Split, Try To Work It Out: Bypassing Conflicts in Multi-Agent Pathfinding.
\newblock \emph{Proceedings of the International Conference on Automated Planning and Scheduling}, 25(1): 47--51.

\bibitem[{Damani et~al.(2021)Damani, Luo, Wenzel, and Sartoretti}]{damani2021primal2}
Damani, M.; Luo, Z.; Wenzel, E.; and Sartoretti, G. 2021.
\newblock PRIMAL $ \_2 $: Pathfinding via reinforcement and imitation multi-agent learning-lifelong.
\newblock \emph{IEEE Robotics and Automation Letters}, 6(2): 2666--2673.

\bibitem[{Gama et~al.(2019)Gama, Marques, Leus, and Ribeiro}]{graph_nn_2017}
Gama, F.; Marques, A.~G.; Leus, G.; and Ribeiro, A. 2019.
\newblock Convolutional Neural Network Architectures for Signals Supported on Graphs.
\newblock \emph{IEEE Transactions on Signal Processing}, 67(4): 1034–1049.

\bibitem[{Li et~al.(2022)Li, Chen, Harabor, Stuckey, and Koenig}]{li2022mapf-lns2}
Li, J.; Chen, Z.; Harabor, D.; Stuckey, P.~J.; and Koenig, S. 2022.
\newblock MAPF-LNS2: Fast Repairing for Multi-Agent Path Finding via Large Neighborhood Search.
\newblock \emph{Proceedings of the AAAI Conference on Artificial Intelligence}, 36(9): 10256--10265.

\bibitem[{Li et~al.(2019)Li, Felner, Boyarski, Ma, and Koenig}]{wdgLi2019}
Li, J.; Felner, A.; Boyarski, E.; Ma, H.; and Koenig, S. 2019.
\newblock Improved Heuristics for Multi-Agent Path Finding with Conflict-Based Search.
\newblock In \emph{Proceedings of the Twenty-Eighth International Joint Conference on Artificial Intelligence, {IJCAI-19}}, 442--449. International Joint Conferences on Artificial Intelligence Organization.

\bibitem[{Li, Ruml, and Koenig(2021)}]{li2021eecbs}
Li, J.; Ruml, W.; and Koenig, S. 2021.
\newblock EECBS: A bounded-suboptimal search for multi-agent path finding.
\newblock In \emph{Proceedings of the AAAI Conference on Artificial Intelligence (AAAI)}, 12353--12362.

\bibitem[{Li et~al.(2020{\natexlab{a}})Li, Tinka, Kiesel, Durham, Kumar, and Koenig}]{rhcrLi2020}
Li, J.; Tinka, A.; Kiesel, S.; Durham, J.~W.; Kumar, T. K.~S.; and Koenig, S. 2020{\natexlab{a}}.
\newblock Lifelong Multi-Agent Path Finding in Large-Scale Warehouses.
\newblock In \emph{Proceedings of the 19th International Conference on Autonomous Agents and MultiAgent Systems}, AAMAS '20, 1898–1900. Richland, SC: International Foundation for Autonomous Agents and Multiagent Systems.

\bibitem[{Li et~al.(2020{\natexlab{b}})Li, Gama, Ribeiro, and Prorok}]{li2020gnn}
Li, Q.; Gama, F.; Ribeiro, A.; and Prorok, A. 2020{\natexlab{b}}.
\newblock Graph Neural Networks for Decentralized Multi-Robot Path Planning.
\newblock arXiv:1912.06095.

\bibitem[{Li et~al.(2021)Li, Lin, Liu, and Prorok}]{li2021magat}
Li, Q.; Lin, W.; Liu, Z.; and Prorok, A. 2021.
\newblock Message-Aware Graph Attention Networks for Large-Scale Multi-Robot Path Planning.
\newblock arXiv:2011.13219.

\bibitem[{Okumura(2022)}]{okumura2022lacam}
Okumura, K. 2022.
\newblock LaCAM: Search-Based Algorithm for Quick Multi-Agent Pathfinding.
\newblock arXiv:2211.13432.

\bibitem[{Okumura et~al.(2022)Okumura, Machida, Défago, and Tamura}]{pibt}
Okumura, K.; Machida, M.; Défago, X.; and Tamura, Y. 2022.
\newblock Priority inheritance with backtracking for iterative multi-agent path finding.
\newblock \emph{Artificial Intelligence}, 310: 103752.

\bibitem[{Sartoretti et~al.(2019)Sartoretti, Kerr, Shi, Wagner, Kumar, Koenig, and Choset}]{sartoretti2019primal}
Sartoretti, G.; Kerr, J.; Shi, Y.; Wagner, G.; Kumar, T.~S.; Koenig, S.; and Choset, H. 2019.
\newblock Primal: Pathfinding via reinforcement and imitation multi-agent learning.
\newblock \emph{IEEE Robotics and Automation Letters}, 4(3): 2378--2385.

\bibitem[{Sharon et~al.(2015)Sharon, Stern, Felner, and Sturtevant}]{sharon2015cbs}
Sharon, G.; Stern, R.; Felner, A.; and Sturtevant, N.~R. 2015.
\newblock Conflict-based search for optimal multi-agent pathfinding.
\newblock \emph{Artificial Intelligence}, 219: 40--66.

\bibitem[{Sharon et~al.(2013)Sharon, Stern, Goldenberg, and Felner}]{ictsMDDSharon2013}
Sharon, G.; Stern, R.; Goldenberg, M.; and Felner, A. 2013.
\newblock The increasing cost tree search for optimal multi-agent pathfinding.
\newblock \emph{Artificial Intelligence}, 195: 470--495.

\bibitem[{Stern et~al.(2019)Stern, Sturtevant, Felner, Koenig, Ma, Walker, Li, Atzmon, Cohen, Kumar, Boyarski, and Bartak}]{stern2019mapfbenchmark}
Stern, R.; Sturtevant, N.~R.; Felner, A.; Koenig, S.; Ma, H.; Walker, T.~T.; Li, J.; Atzmon, D.; Cohen, L.; Kumar, T. K.~S.; Boyarski, E.; and Bartak, R. 2019.
\newblock Multi-Agent Pathfinding: Definitions, Variants, and Benchmarks.
\newblock \emph{Symposium on Combinatorial Search (SoCS)}, 151--158.

\bibitem[{Wang et~al.(2023)Wang, Xiang, Huang, and Sartoretti}]{wang2023scrimp}
Wang, Y.; Xiang, B.; Huang, S.; and Sartoretti, G. 2023.
\newblock SCRIMP: Scalable Communication for Reinforcement- and Imitation-Learning-Based Multi-Agent Pathfinding.
\newblock arXiv:2303.00605.

\end{thebibliography}

\clearpage
\appendix

\setcounter{figure}{0}
\renewcommand{\thefigure}{A\arabic{figure}}
\setcounter{table}{0}
\renewcommand{\thetable}{A\arabic{table}}


\section{Additional Analysis of CS-PIBT }


\subsection{Quantitative Analysis}
\begin{figure}[h!]
    \centering
    \begin{subfigure}{0.4\textwidth}
         \centering
         \includegraphics[width=\textwidth]{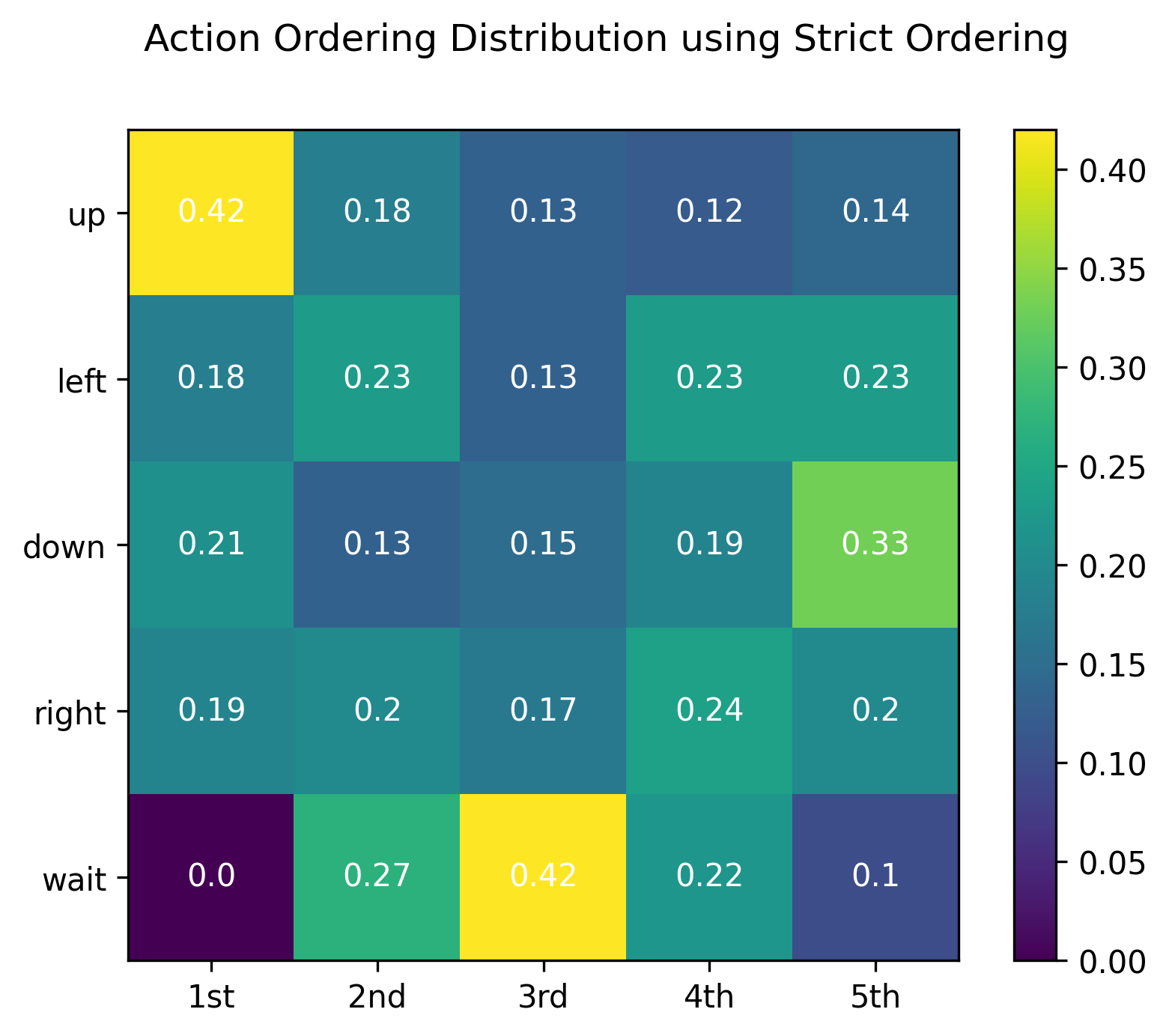}
         \subcaption{The value of cell [action, $k$] is the probability that action is the $k^{th}$ preference when converting probability distributions to action orderings using strict ordering.}
         \label{fig:ordering-strict}
    \end{subfigure}
    \hfill
    \begin{subfigure}{0.4\textwidth}
         \centering
         \includegraphics[width=\textwidth]{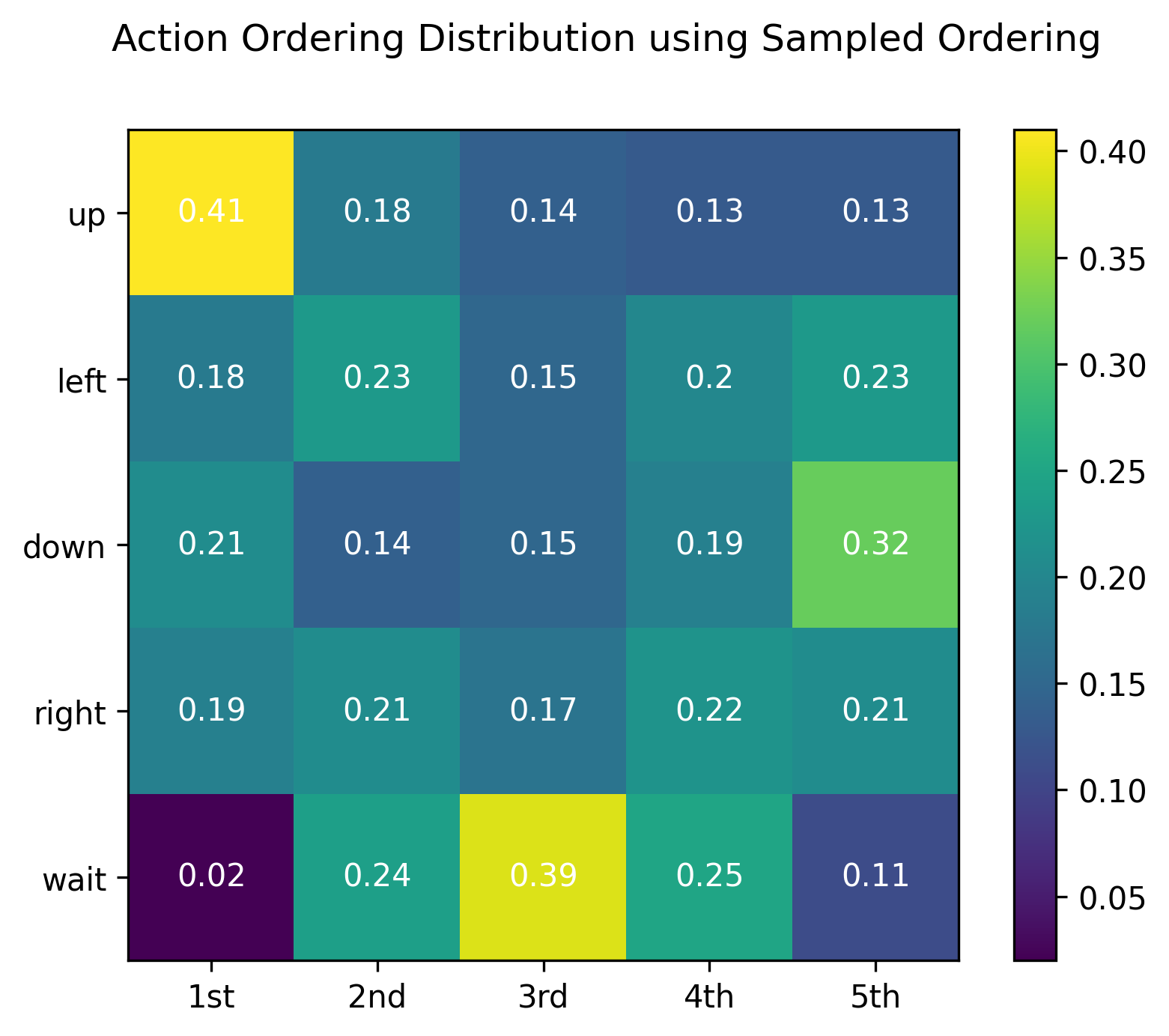}
         \subcaption{The value of cell [action, $k$] is the probability that action is the $k^{th}$ preference when converting probability distributions to action ordering using sampled ordering.}
         \label{fig:ordering-sampled}
    \end{subfigure}
    \caption{Action ordering distributions}
    \label{fig:sampled-vs-ordering}
\end{figure}

Converting the probability distribution to an action ordering using sampling was substantially better than using a strict ordering of probabilities. One possibly hypothesis is that the distributions induced are different as the strict ordering could over-prefer higher probability actions. Figure \ref{fig:sampled-vs-ordering} visualizes the action distribution when converting probabilities into action orders using strict ordering and sampling for agents not at their goal state. We notice they are very similar, indicating this is not the cause of the performance difference within CS-PIBT. This specific distribution is for 50 agents not at their goal state. 
Note agents at their goal had different overall distributions (i.e. ``wait" probabilities were 1st nearly all the time), but strict and sampled ordering produced similar representative distributions there too.


\subsection{Qualitative Analysis}
We provide \url{https://drive.google.com/drive/folders/1iVSXbMHINnCsMzMPVi4k7P99RZFX6qbQ} with visualization animations of MAGAT with CS-Naive, CS-PIBT strict ordering (without randomness), and CS-PIBT with sampling.
Obstacles are black and free space is white. Each agent is given a unique color and turns grey when it reaches its goal state.

We observe most failure cases occur due to a few agents getting stuck in deadlock or live-lock due to other agents resting at their nearby goal state. Failure cases with CS-PIBT with strict ordering show live-lock where agents alternate between the same positions, while CS-PIBT with sampling does not show it as often. Thus, sampling is effective in reducing these instances by promoting trying different actions at the same state. We also note that with CS-Naive, agents in crowds take quite long (or possibly never) to leave/travel through the crowd, while with CS-PIBT they are able to travel effectively through them.

\end{document}